\documentclass[12pt,a4paper]{article}
\usepackage{fullpage}
\usepackage[centertags]{amsmath}
\usepackage{amsbsy}
\usepackage{amsfonts}
\usepackage{amssymb}
\usepackage[dvips]{graphicx}
\usepackage[dvips,monochrome]{color}
\allowdisplaybreaks[1]
\newcommand{\lsim}{\mathrel{\mathop{\kern 0pt \rlap
  {\raise.2ex\hbox{$<$}}}
  \lower.9ex\hbox{\kern-.190em $\sim$}}}
\newcommand{\gsim}{\mathrel{\mathop{\kern 0pt \rlap
  {\raise.2ex\hbox{$>$}}}
  \lower.9ex\hbox{\kern-.190em $\sim$}}}

\def\@fnsymbol#1{if case#1\hbox{}\or*\or\dagger\or\ddagger\or\mathcar''278\or\mathchar''27B\or|\or**\or\dagger\dagger\or\ddagger\ddagger\else\@ctrerr\fi\relax}
\long\def\symbolfootnote[#1]#2{\begingroup%
\def\thefootnote{\fnsymbol{footnote}}\footnote[#1]{#2}\endgroup}

\long\def\letterfootnote[#1]#2{\begingroup%
\def\thefootnote{\alph{footnote}}\footnote[#1]{#2}\endgroup}

\begin{document}

\begin{center}
\large\bfseries
Heavy-ion collisions described by a new QMD code interfaced to FLUKA: model
validation by comparisons with experimental data concerning neutron and
charged fragment production 
\\[0.5cm]
\normalsize\normalfont
M.V. Garzelli$^{1,2,\,}$\symbolfootnote[4]{$\,$ {\it Corresponding author,  
$\,$e-mail:} garzelli@mi.infn.it}, 
F. Ballarini$^{3,4}$, G. Battistoni$^2$, 
F. Cerutti$^{5}$, A. Fass\`o$^6$,
A. Ferrari$^5$, E.~Gadioli$^{1,2}$, A.~Ottolenghi$^{3,4}$, L.S.~Pinsky$^7$, J.~Ranft$^8$, P.R.~Sala$^2$
\\[0.5cm]
\small\itshape
$^{1}\it{University}$ of Milano, Physics Department, via Celoria 16, I-20133, Milano, Italy
\\
$^{2}\it{INFN}$ Milano, via Celoria 16, 
I-20133, Milano, Italy \\
$^{3}\it{University}$ of Pavia, Nuclear and Theoretical Physics Department,\\
via Bassi 6, I-27100, Pavia, Italy \\
$^{4}\it{INFN}$ Pavia, via Bassi 6, I-27100, Pavia, Italy \\
$^{5}\it{CERN}$, CH-1211, Geneva, Switzerland \\
$^{6}\it{SLAC}$, Stanford, US \\
$^{7}\it{University}$ of Houston, Department of Physics, Houston, US \\
$^{8}\it{Siegen}$ University, Siegen, Germany \\
\small\itshape
\small\itshape
\end{center}
\begin{abstract}
A new code, based on the Quantum Molecular Dynamics theoretical approach,
has been developed and
interfaced to the FLUKA evaporation/fission/Fermi break-up module. 
At present, this code is undergoing a series of validation tests.
In this paper its predictions are  compared to measured
charged fragment yields and double differential
neutron spectra in thin target  heavy-ion reactions, 
at bombarding energies of about 100 MeV/A. The comparisons with the predictions
of a modified version of RQMD 2.4 originally developed 
in Frankfurt \cite{sorge},
already available in the FLUKA code, are presented and potential improvements 
are briefly sketched.
\end{abstract}


\section{Introduction}
\label{Introduction}
$\,\,\,\,\,\,\,$ Nucleon-nucleus interactions and heavy-ion collisions 
at non relativistic bombarding energies can be described by macroscopic
models or, at a more fundamental level, by microscopic models considering the
nucleonic degrees of freedom.

The experimental information shows that {\it nucleon-nucleus} 
collisions are dominated both by mean field effects 
and by short-range correlations due to 
nucleon-nucleon forces at the shortest distances. 
Models based on the Intra Nuclear Cascade (INC) theory can be used
to simulate these events, since they incorporate both these aspects:
the projectile hadron moves along a curved trajectory 
in the target nucleus mean field
generated by Coulomb and nuclear forces.
When the projectile hits one of the target nucleons along its path, 
a scattering can occur,
which leads to an abrupt change in the trajectory of both particles.
As a first approximation, the potential originated by the incident
nucleon can be neglected with respect to the one originated by the target 
ion, especially for heavy targets. 

At the contrary, when {\it nucleus-nucleus} 
collisions are considered, each nucleus ge\-ne\-ra\-tes
a potential field that affects the movement of the other. In other
words, each nucleon of the interacting nuclei feels the field due to
both the nucleons of its same nucleus, and to the nucleons of the interacting
partner. Thus, the force that each nucleon globally experiences is given by a
superposition of nucleon-nucleon interactions.

Quantum Molecular Dynamics (QMD) models allow to calculate this force
and to study the dynamical evolution of a nucleus-nucleus collision. 
In these models the Hamiltonian includes an effective interaction
part depending on two-nucleon forces. Many-body effects can also be
incorporated. A scattering term can be included as well, as in INC 
codes.

Different research groups have developed models based on
QMD theory, which mainly differ in the expression of the Hamiltonian.
In the following of this paper the model developed by our group is
briefly described (see also Ref.~\cite{paviamv}). A few examples of the 
model predictions and their comparisons with those 
of a modified version~\cite{vecchiavarenna} of 
the RQMD 2.4 code, originally
developed by H. Sorge at Frankfurt~\cite{sorge}, 
are presented in the following sections. The modifications were made
with  the aim of evaluating the yields and the double differential spectra
of the fragments which may by produced in the interaction, since
the original version gave only the emitted nucleon spectra. 
Both the new QMD
code and the RQMD code 
have been interfaced to the evaporation/fission/Fermi break-up
module available in the FLUKA Monte Carlo transport and interaction
code~\cite{fluka}, which simulates the de-excitation of the hot fragments 
produced at the end of the fast stage of heavy-ion collisions. 
Their de-excitation indeed may occur on a time scale 
far larger than that of the primary ion-ion collision, 
and is thus better described by different models, based on statistical
instead of dynamical considerations.  

\section{Theoretical framework} 
\begin{figure}[t!]
\begin{center}
\includegraphics[bb=51 51 405 300,  width=0.80\textwidth]{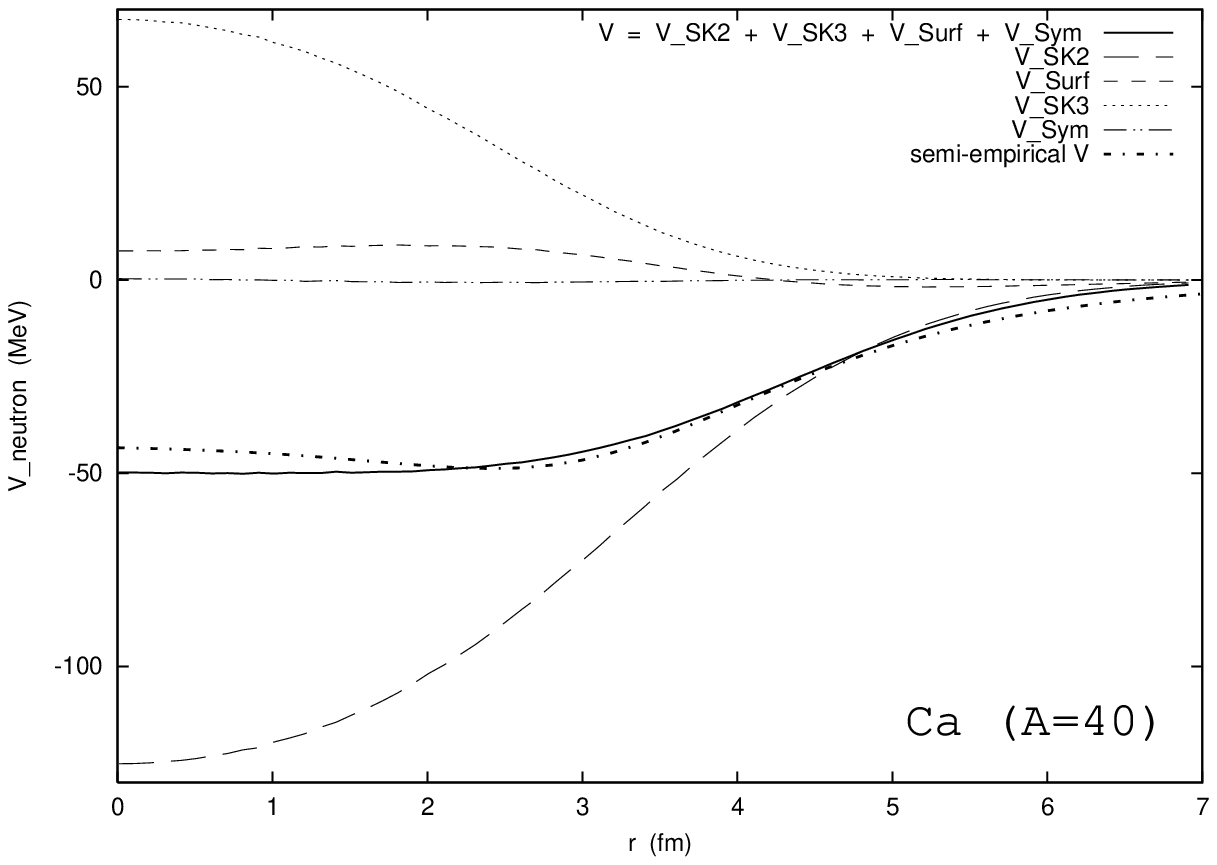}
\includegraphics[bb=51 51 405 300,  width=0.80\textwidth]{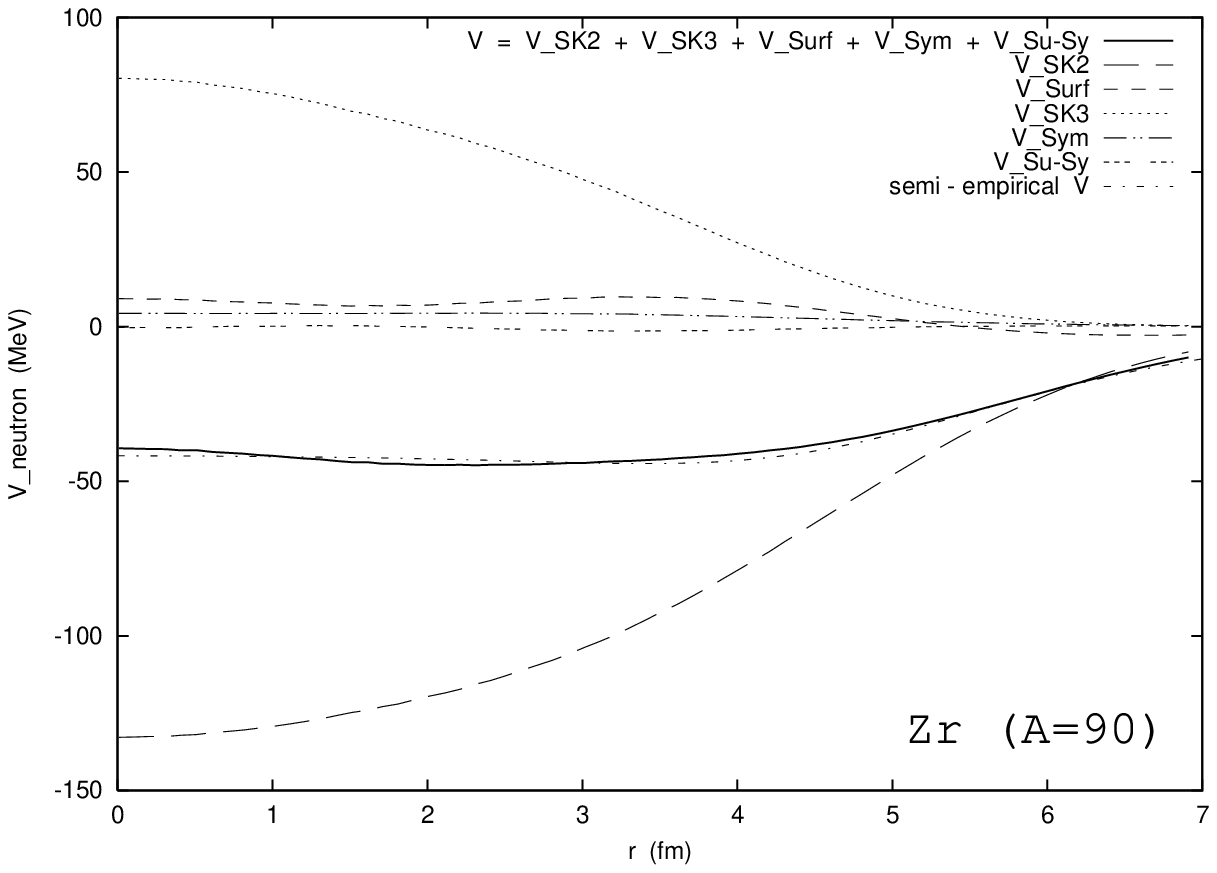}
\caption{Radial dependence of the  neutron potential for
$^{40}\mathrm{Ca}$ nuclei ({\it top}) 
and for 
$^{90}\mathrm{Zr}$ nuclei ({\it bottom}), including
the {\it Skyrme-II} attractive term (long-dashed line), the 
{\it Skyrme-III} repulsive
term (dotted line), the {\it symmetry} term (dashed-double-dotted line)
 and the total (attractive + repulsive)
{\it surface} term (short-dashed line). 
The sum of all these terms is given
by the solid line and is compared to the semi-empirical 
central component of the neutron potential calculated
according to the parameter set OB of ~Ref.~\cite{koura} (dot-dashed line).}
\label{potmed}
\end{center}
\end{figure}

In QMD models each nucleon is described in coordinate space as a gaussian 
wave-packet,
\begin{equation}
\phi_i(\vec{r},t)=\frac{1}{(2\pi\sigma^2_r)^{3/4}}e^{-\frac{(
\vec{r}-\vec{R_i}(t))^2}{4\sigma^2_r}}e^{i \frac{\vec{r}\cdot\vec{P_i}(t)}{\hbar}}\, . 
\label{wave}
\end{equation}
$\vec{R_i}$ and $\vec{P_i}$ are the average 
values of the position
and the momentum of the $i-th$ nucleon, $\sigma_r$ is the gaussian space 
width related to the nucleon momentum distribution width $\sigma_p$ by the
uncertainty relationship $\sigma_r \sigma_p = \hbar/2$. For a nucleus
with mass number $A$,
the nuclear wave-function is approximated by the product of $A$ nucleon 
wave-functions:
\begin{equation}
\phi(\vec{r}_1,\vec{r}_2,......\vec{r}_A, t) 
=\prod_{i=1}^A \phi_i(\vec{r}_i,t)
\, ,
\end{equation}
neglecting the fermionic nature of the nucleons. 
In an approximate way, their fermionic behaviour 
is taken into account a) in the initialisation scheme, when 
nucleon spatial coordinates and momenta are sampled, according to the 
ex\-pe\-ri\-men\-tal\-ly 
observed density distributions (see next section), 
b) by including Pauli blocking factors 
forbidding  the collisions 
which lead a nucleon to a phase-space region already filled.
The Hamiltonian operator is a sum of a kinetic and an
interaction term:
\begin{equation}
\hat{H}=\hat{T}+\hat{H}_{int} \, .
\end{equation}
The interaction part of the Hamiltonian
is the sum of many effective terms:
\begin{equation}
\hat{H}_{int}=\hat{H}_{int,\, eff}=\hat{H}_{Skyrme-II}+\hat{H}_{Skyrme-III}+
\hat{H}_{symmetry} + \hat{H}_{surface} + \hat{H}_{Coulomb} \, .
\label{hamilt}
\end{equation}
Here, the Skyrme-II two-body interaction term is attractive, while
the Skyrme-III three-body interaction term mimics repulsive forces, that are 
crucial
for reproducing the nuclear matter sa\-tu\-ra\-tion properties at normal density. 
The symmetry term takes into account the dependence of nucleon forces
on the isospin. 
The surface term leads to two contributions, one repulsive
and the other attractive: their sum reproduces
the decrease of the potential at low nuclear radii, 
especially for low mass nuclei.  
Examples of the relative importance
of these contributions to the total average central
potential for neutrons in
$^{40}\mathrm{Ca}$ ({\it top} panel) and $^{90}\mathrm{Zr}$ 
({\it bottom} panel) are shown in Fig.~\ref{potmed}. 
These potentials are compared to those proposed by~\cite{koura}, which are
obtained by a totally different approach.
At last, a Coulomb term is added to those of eq.~(\ref{hamilt}), 
to take into account the $p-p$ Coulomb repulsion.

Evolution of nuclear states is determined by the Hamiltonian. To evaluate
nucleon-nucleon scattering we use {\it n-p} and {\it p-p} free
cross sections which, as a first approximation, we assume to be isotropic.
Inclusion of the angular distributions of scattering cross-section, as well of
inelastic channels leading to pion and other 
particle production via resonance formation, represents
an important future extension of our model, especially when approaching
relativistic energies.     
Neutrons and protons are distinguished 
whenever possible in our code. In particular, the symmetry and Coulomb term
of the Hamiltonian, as well as the nucleon-nucleon scattering term,
take into account the different behaviour of these particles.
As an example, there is a factor about three between the free {\it n-p} and
{\it p-p }
cross-sections
at the energies relevant for this work, as follows from isospin and
charge considerations.  

\section{Nuclear ground states} 
Nuclear states are initialized by sampling initial spatial
coordinates and momenta for all nucleons in each nucleus.  
Nucleon-nucleon 
phase-space distances should be large enough to satisfy Pauli
Priciple, leading to smooth nuclear density and momentum profiles. 
Peculiar of our calculations is the fact that
the total energy of each ion, calculated as a sum of the kinetic
and the interaction energy, according to the Hamiltonian
introduced in previous section, is exactly
constrained by the experimentally
observed nuclear bin\-ding energy.
The difference between $n$ and $p$ free masses is also taken into
account and attention is paid to ensure conservation of energy and momentum.
The evolution of each nuclear state according to the Hamiltonian
is followed
for a time-scale of 250 - 300 fm/c. If, in this time interval, 
no spurious emission of nucleons 
occurs and root mean square radius oscillations 
maintain within a few percent,
the configuration is accepted and stored off-line. It can then be used
to simulate ion-ion (or nucleon-ion) collision events.

\section{Ion - ion collisions}
The nuclear states stored during the initialisation procedure
can be used to si\-mu\-la\-te ion-ion collisions. To this purpose, 
for any given impact parameter value, the ions are boosted one towards the 
other to mimic
the collision process. 
As soon as colliding nuclei come close to each other, nuclear
forces due to the target (projectile) nucleons become 
increasingly important in
affecting the projectile (target) nucleon behaviour. 
Additionally, Coulomb 
interaction, because of its infinite range, appears to be crucial
since the beginning of the simulation, also 
when the ions are still far from each other, especially for determining
the distance between the two ion centers when the ions come in contact, 
which slightly differs from the impact parameter.
When the incoming nuclei overlap, an hot excited system, characterized by an 
increased density and pressure, is created. 
The compression phase is followed by an expansion stage,
in\-vol\-ving density fluctuations, 
leading to the formation of excited fragments. During this fast stage, to
define a temperature can be a very difficult or even an impossible task, 
especially at energies of several tens MeV/$A$.
At lower energies, repeated
nucleon-nucleon collisions gradually lead to a thermalization
of the biggest excited systems (pre-equilibrium phase), followed by
isotropic emission.  
On the other hand, at higher energies, the expansion phase starts before 
this thermalization process could be completed. 

Hot fragments are
defined on the basis of nucleon final configuration, 
without any re\-fe\-ren\-ce 
to their target or projectile initial belonging. 
Their de-excitation involves the competition of different processes whose
probability of occurrence depends on the nuclear species 
and the energy. The processes considered in our calculations and simulated by
the de-excitation module of the FLUKA code,
account for $\gamma$ emission, evaporation,
Fermi break-up and fission. Fermi break-up is used for excited 
light nuclei and may produce up to six fragments. The FLUKA evaporation 
routine allows the emission 
of particle and light fragments, up to a mass number $A=24$.
Further details on the FLUKA de-excitation module
can be found in Ref.~\cite{flukacern}.

\section{Results: double - differential neutron spectra and
fragment yields}
\begin{figure}[t!]
\includegraphics[bb=50 50 680 460, width=0.80\textwidth]{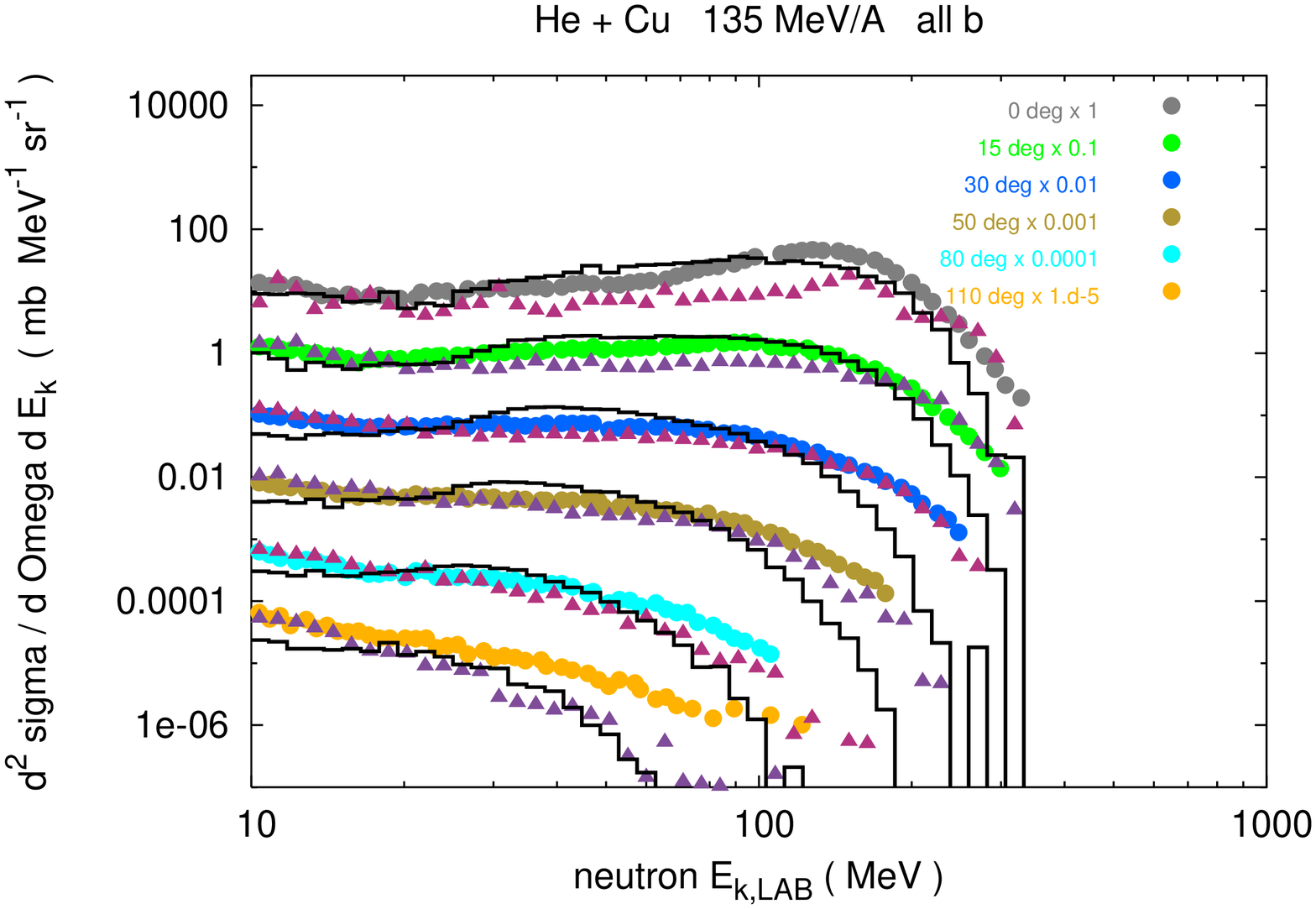}
\\
\\
\\
\includegraphics[bb=50 50 680 460, width=0.80\textwidth]{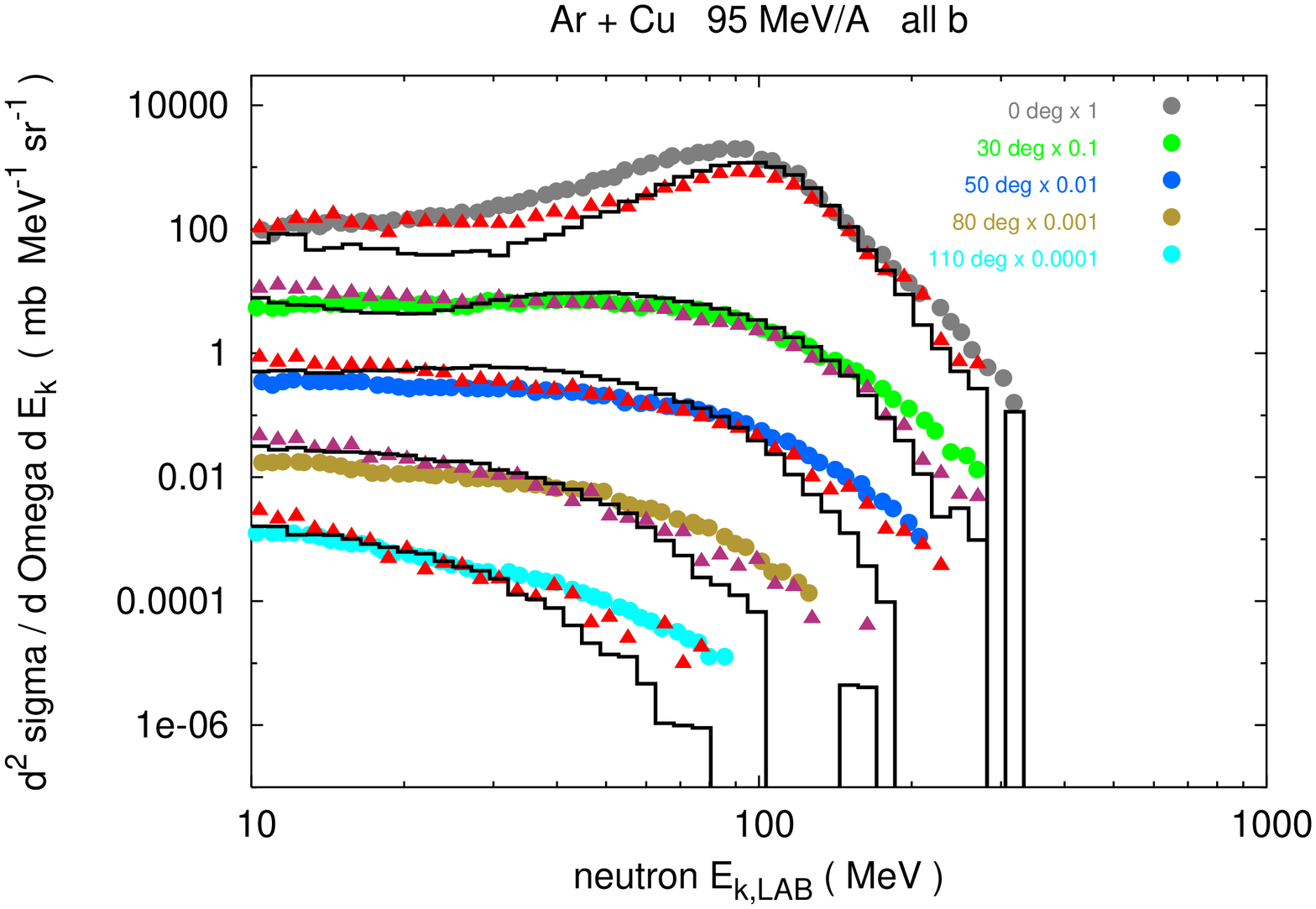}

\caption{Double-differential neutron spectra
for He + Cu at 135 MeV/$A$ bombarding
energy (upper figure) and Ar + Cu at 95 MeV/$A$ (lower figure).
In each figure the spectra are given as a function of the neutron energy, and,
from top to bottom, each spectrum corresponds to emission angles $\theta$  = 0$^\mathrm{o}$, 15$^\mathrm{o}$, 
30$^\mathrm{o}$, 50$^\mathrm{o}$, 80$^\mathrm{o}$, 110$^\mathrm{o}$. 
The theoretical distributions 
predicted by QMD~+~FLUKA (filled triangles) and by RQMD~+~FLUKA (histograms)
are compared to the experimental data of Ref.~\cite{chiba} (filled
circles)}. 
\label{figurecross}
\end{figure}

\begin{figure}[t!]
\includegraphics[bb=0 0 680 460, width=0.80\textwidth]{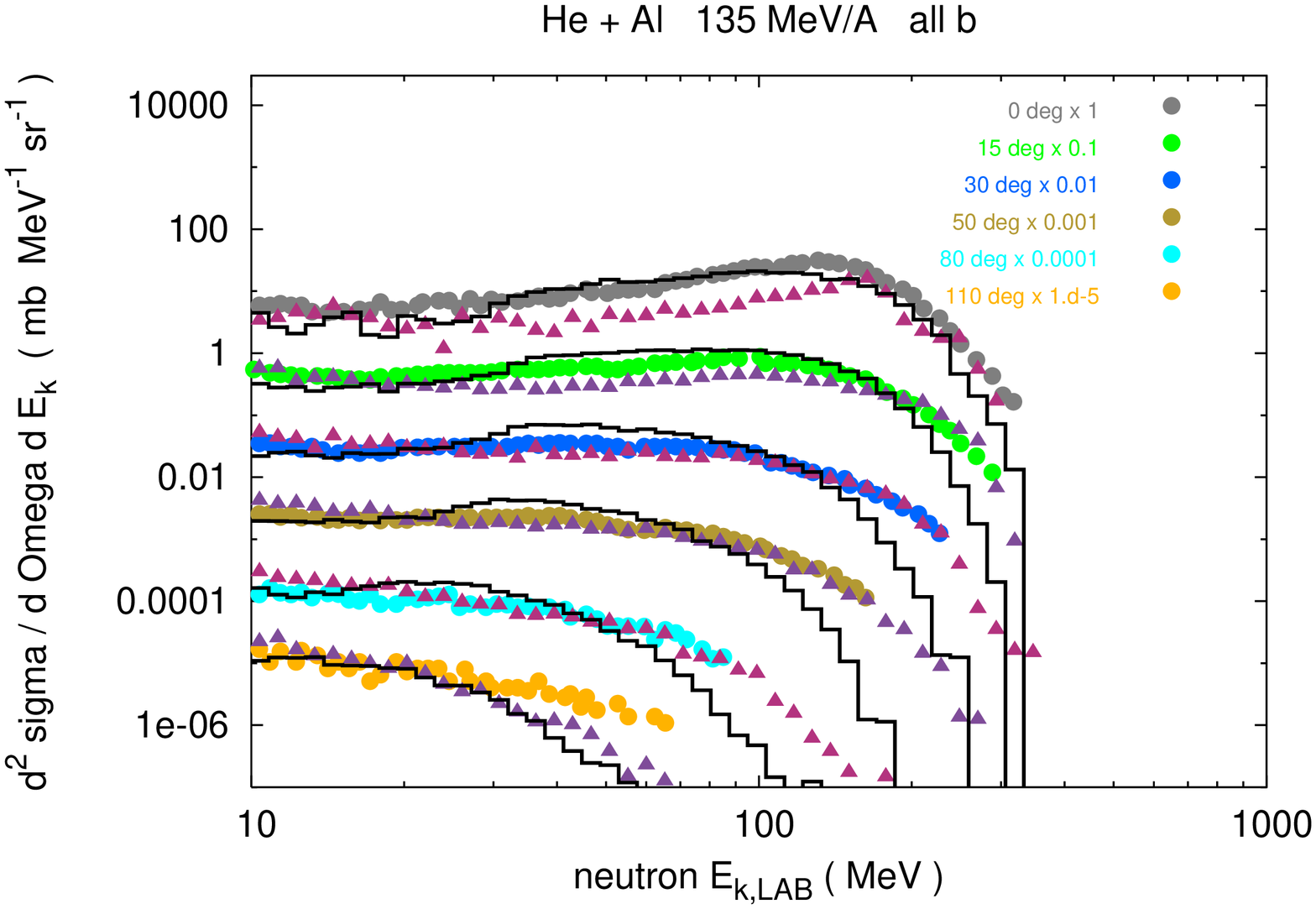}
\\
\includegraphics[bb=0 0 680 460, width=0.80\textwidth]{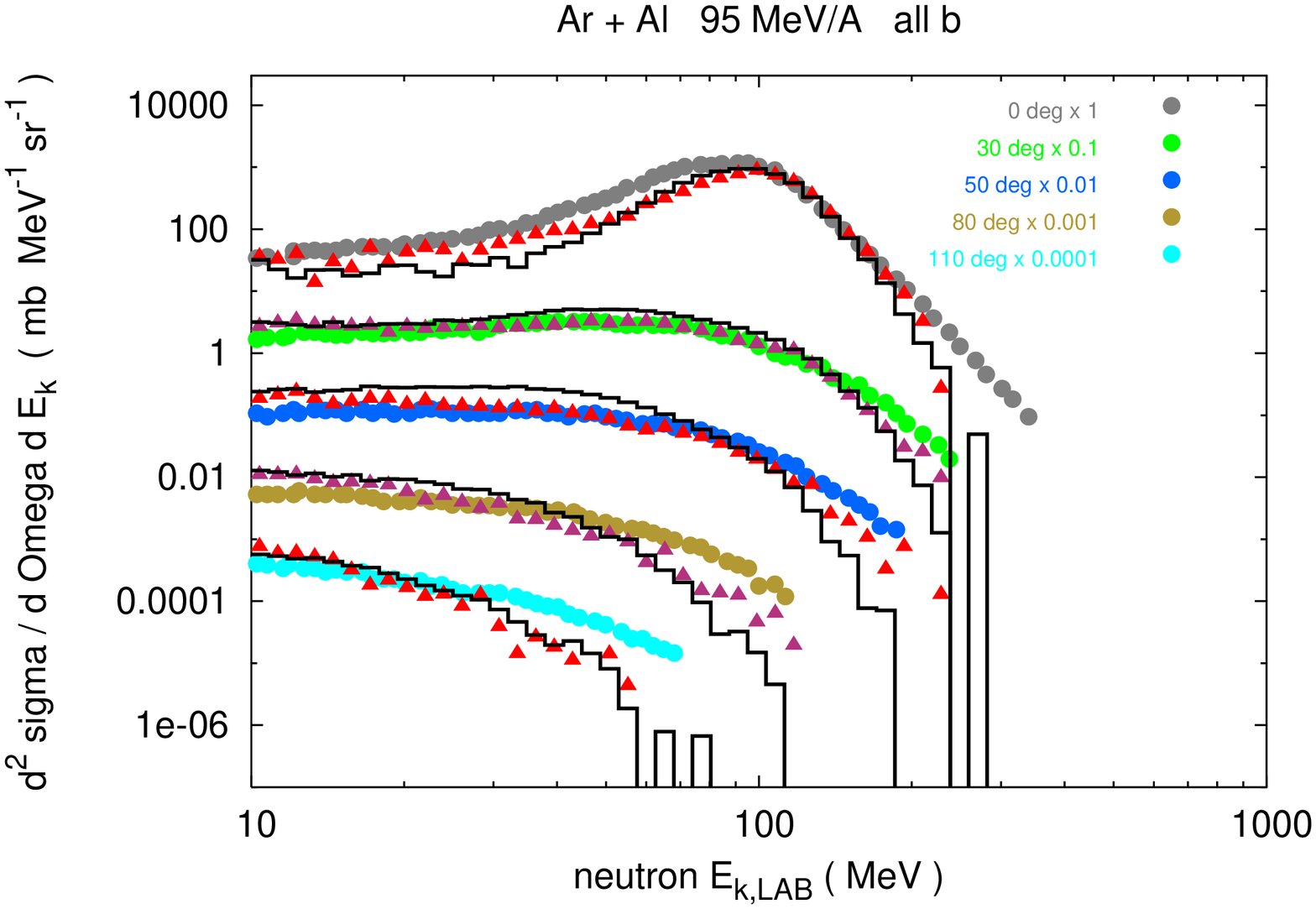}

\caption{Double-differential neutron spectra
for He + Al at 135 MeV/$A$ bombarding
energy (upper figure) and Ar + Al at 95 MeV/$A$ (lower figure).
In each figure the spectra are given as a function of the neutron energy, and,
from top to bottom, each spectrum corresponds to emission angles $\theta$  = 0$^\mathrm{o}$, 15$^\mathrm{o}$, 
30$^\mathrm{o}$, 50$^\mathrm{o}$, 80$^\mathrm{o}$, 110$^\mathrm{o}$.
The theoretical distributions 
predicted by QMD~+~FLUKA (filled triangles) and by RQMD~+~FLUKA (histograms)
are compared to experimental data of Ref.~\cite{chiba} (filled
circles).}
\label{figurecross2}
\end{figure}

We have analyzed the experimental data of Ref.~\cite{chiba}.
These authors in an thin target ex\-pe\-ri\-ment 
measured the double-differential spectra of neutrons emitted in the interaction
of He, C, Ne ions, at 135 MeV/$A$ bombarding energy, and of Ar ions,
at 95 MeV/$A$ bombarding energy, with C, Al, Cu and Pb ions.
Examples of the comparison of the experimental data and our theoretical 
predictions are shown in Fig.~\ref{figurecross} 
and Fig.~\ref{figurecross2} for the reactions induced on the aluminium and
copper targets.
Different sets of data on the same plot refer to different emission
angles. The absolute yields have been 
multiplied 
by decreasing scaling factors (negative powers of 10) 
for increasing emission angle.  
As already mentioned at the beginning of this paper, in our analysis we used
both the QMD code and the 
modified version of the RQMD 2.4 code~\cite{vecchiavarenna}.

The two codes differ quite substantially. The Hamiltonian included in the 
Sorge code is fully
relativistic, but involves only Skyrme-II and Skyrme-III interactions.
No di\-stinc\-tion is made between neutrons and protons and Coulomb effects
are not included. On the other hand, the Hamiltonian in our QMD code
can implement relativistic or non-relativistic kinematics, according
to a switch, but only instantaneous interactions are included 
(non-relativistic potential). 
As already mentioned, neutrons and protons are 
distinguished both considering their mass and their isospin.   
In the figures, 
the results of the si\-mu\-la\-tions performed with our QMD code coupled to 
FLUKA are given by triangles, 
while those of the simulations performed with RQMD + FLUKA
are represented by  histograms. No nor\-ma\-li\-za\-tion factor has been included in 
the simulations. 

In general, both models show an overall good agreement 
with the experimental data. At a more detailed level,
the figures show that our QMD is capable
of predicting better than RQMD the behaviour of  neutron spectra at
intermediate emission angles (30$^\mathrm{o}$ $-$ 80$^\mathrm{o}$), 
especially at the highest energy end of the spectra, while it underestimates the 
emissions in the very forward direction. Finally, both models show 
a comparable agreement with the experimental 
data at backward angles ($\ge$~90$^\mathrm{o}$).
We expect that the disagreement between data and QMD calculations in 
the forward and backward directions can be reduced by implementing 
in the code the angular dependence of {\it n}-{\it p} cross sections, which
should increase the yield of particles emitted
back and forth along the collision axis. 

\begin{figure}[b!]
\includegraphics[bb=51 51 720 500, width=0.80\textwidth]{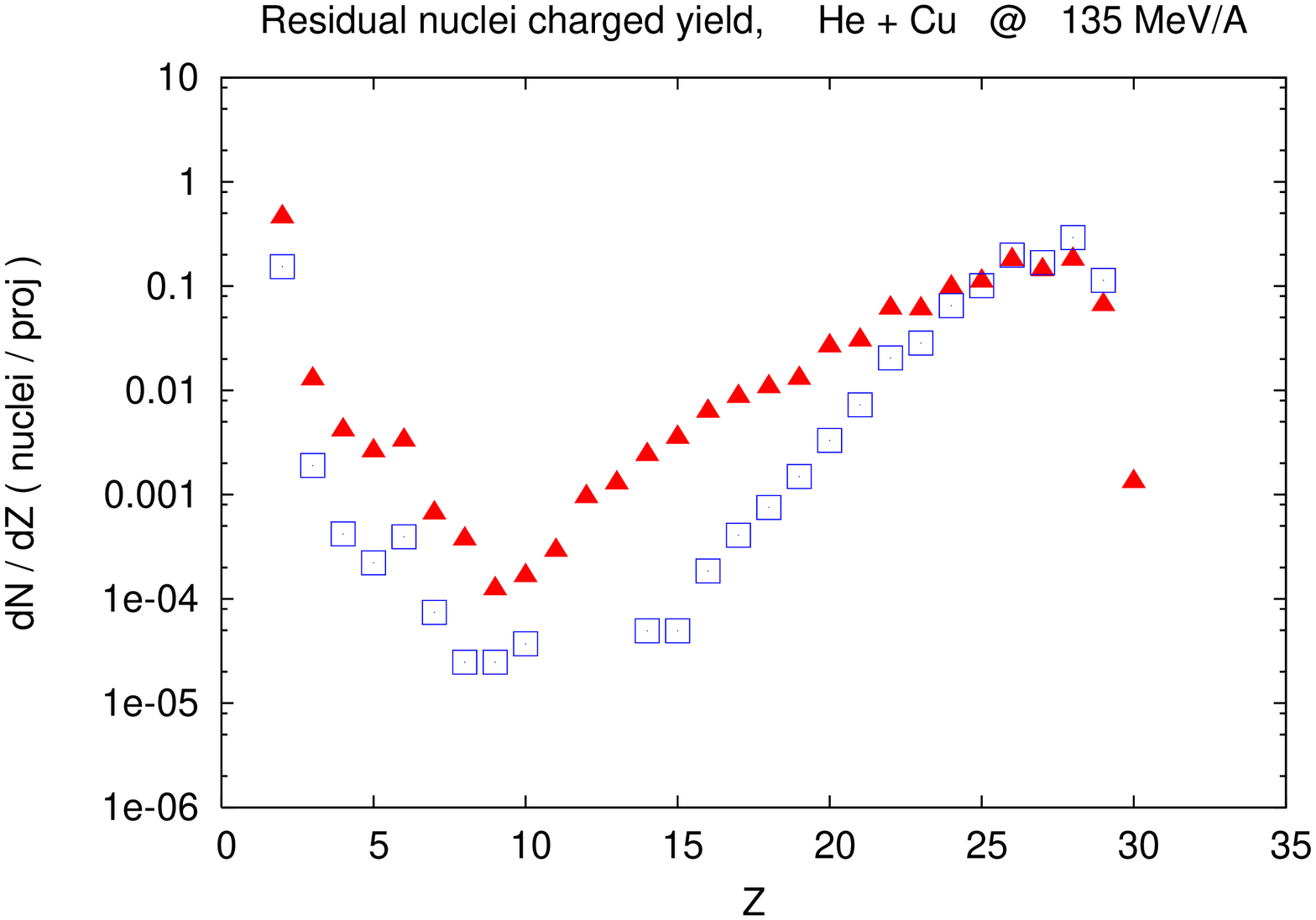}

\includegraphics[bb=51 51 720 500, width=0.80\textwidth]{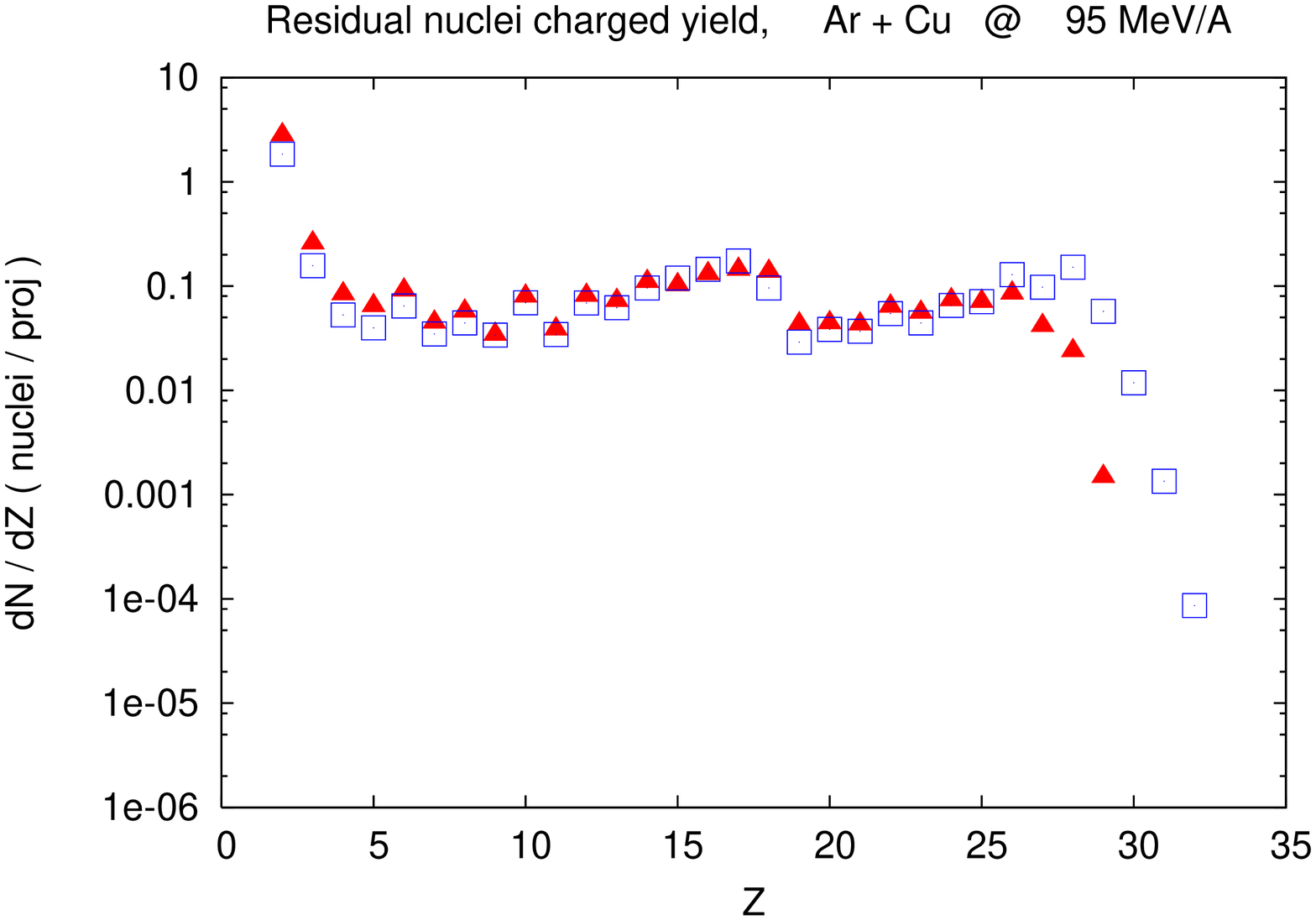}
\caption{Residual nuclei charged yield for He + Cu at 135 MeV/$A$ bombarding
energy ($\eta = 0.88$) ({\it top}) and for Ar + Cu at 95 MeV/$A$ ($\eta = 0.22$) ({\it bottom}), as
predicted by QMD~+~FLUKA (filled triangles) and by RQMD~+~FLUKA (boxes). 
}
\label{heraarrafrag}
\end{figure}

We have also calculated the yields of fragments emitted in the
same reactions. 
Unfortunately, this information is not experimentally available, thus, we only 
compare the predictions of the two codes. Let us discuss the results
concerning the fragment charge yield distributions. These are shown
in the upper and in the lower panels of Fig.~\ref{heraarrafrag}
for the He~+~Cu and Ar + Cu interactions, respectively.
We find that the QMD code predicts an higher He
fragment yield and the di\-scre\-pan\-cies between the two model predictions
increase with increasing reaction mass asymmetry 
\begin{equation}
\eta = \frac{|A_{proi}-A_{targ}|}{A_{proi}+A_{targ}} \,\, .
\end{equation}  
For reactions characterized by a large mass asymmetry,
such as He + Cu, the yield of intermediate charge fragments predicted 
by QMD + FLUKA is far  larger that that predicted by 
RQMD + FLUKA as shown in the upper part of Fig. 4. 
These fragments are produced 
as remnants of projectile and target nuclei, in central collisions, and
by evaporation of projectile-like and target-like nuclei, in more peripheral
collisions. 
 The reasons of these discrepancies 
in pre\-dic\-ting such a different number of IMF has still to be understood.
In particular, the way the two codes treat ion-ion 
asymmetric systems, the role of the compressibility and that of the hot 
fragment definition scheme, at the end of the ion overlapping stage, are under
investigation.
 
The description of asymmetric systems by QMD/RQMD models
is more difficult than the description of the symmetric ones, because in the
effective Hamiltonian the light nucleus parameters may be different 
from the heavy nucleus ones as suggested by many authors.
Thus, while in RQMD a fixed parameter set is used for all nuclei,
in QMD we try to incorporate these differences adopting parameters
depending on the nuclear mass, as already done 
in ~\cite{china}. The discrepancy may also be due to the fact that 
QMD Hamiltonian includes a Coulomb term while RQMD does not,
and to the different
way the production of fragments is evaluated in the two codes. In RQMD
calculation only projectile-like and  target-like hot residuals are considered,
while in QMD, as previously reminded, 
hot fragments are
defined on the basis of excited nucleon configurations, 
without any re\-fe\-ren\-ce 
to their target or projectile initial belonging. This may lead to
multifragmentation and to the production of fragments made of nucleons of both 
the projectile and the target.
The discrepancies seem to reduce for smaller
reaction mass asymmetries as shown in the lower part of Fig. 4.

\section{Conclusions}
A code, based on the Quantum Molecular Dynamics approach,
has been developed for describing the fast stage
of heavy-ion collisions at non-relativistic bombarding energies 
and coupled to the FLUKA de-excitation module.
The model is undergoing further improvements, such as the inclusion of pion 
and particle production via resonance formation accor\-ding 
to the isobar model, 
in order to extend its predictions to higher energies. 
The angular dependence of nucleon-nucleon
scattering cross-section needs also to be included.
Comparisons with the experimental data and the predictions 
of other theoretical mo\-dels is encouraging.
In particular, the double-differential spectra of the 
emitted neutrons measured by~\cite{chiba} are quite satisfactorily reproduced.

\end{document}